\RequirePackage[l2tabu,orthodox]{nag}
\documentclass
[11pt,letterpaper]
{article}

\usepackage[dvipsnames]{xcolor}
\PassOptionsToPackage{pagebackref}{hyperref}

\usepackage[notes=true,later=false,camera=false]{dtrt}
\usepackage[utf8]{inputenc}
\usepackage{etex}
\usepackage{ stmaryrd }
\usepackage{xspace,enumerate}
\usepackage[T1]{fontenc}
\usepackage[full]{textcomp}
\usepackage[american]{babel}

\usepackage{mathtools}

\usepackage{amsthm}
\usepackage{amsfonts}
\usepackage{empheq}

   \usepackage{hyperref}
   \hypersetup{
colorlinks=true,
urlcolor=Cerulean,
linkcolor=RoyalBlue,
citecolor=PineGreen,
linktocpage=true,
}

\usepackage{paralist}
\usepackage{turnstile}
\usepackage{mdframed}
\usepackage{tikz}
\usepackage{caption}
\DeclareCaptionType{Algorithm}
\usepackage{newfloat}

\usepackage[capitalise,nameinlink]{cleveref}
\usepackage{thm-restate}

\crefname{lemma}{Lemma}{Lemmas}
\crefname{fact}{Fact}{Facts}
\crefname{theorem}{Theorem}{Theorems}
\crefname{mtheorem}{Theorem}{Theorems}
\crefname{corollary}{Corollary}{Corollaries}
\crefname{claim}{Claim}{Claims}
\crefname{example}{Example}{Examples}
\crefname{algorithm}{Algorithm}{Algorithms}
\crefname{problem}{Problem}{Problems}
\crefname{definition}{Definition}{Definitions}
\crefname{question}{Question}{Questions}
\usepackage{newfloat}
\newtheorem{theorem}{Theorem}[section]
\newtheorem{mtheorem}{Theorem}
\newtheorem*{theorem*}{Theorem}

\newtheorem*{proposition*}{Proposition}

\newtheorem*{lemma*}{Lemma}
\newtheorem{corollary}[theorem]{Corollary}
\newtheorem*{conjecture*}{Conjecture}
\newtheorem{fact}[theorem]{Fact}
\newtheorem*{fact*}{Fact}

\newtheorem*{hypothesis*}{Hypothesis}

\theoremstyle{definition}
\newtheorem{definition}[theorem]{Definition}
\newtheorem*{definition*}{Definition}

\newtheorem{algorithm}[theorem]{Algorithm}

\theoremstyle{remark}

\newtheorem*{claim*}{Claim}
\newtheorem{remark}[theorem]{Remark}
\newtheorem*{remark*}{Remark}

\newtheorem*{observation*}{Observation}

\AtBeginDocument{%

}
\usepackage[
letterpaper,
top=1.2in,
bottom=1.2in,
left=1in,
right=1in]{geometry}
\usepackage{newpxtext} 
  \usepackage[varg,bigdelims]{newpxmath}
\usepackage[scr=rsfso]{mathalfa}
\usepackage{bm} 
\usepackage{microtype}
\usepackage{multirow}

\allowdisplaybreaks
\newcommand{\FormatAuthor}[3]{
\begin{tabular}{c}
#1 \\ {\small\texttt{#2}} \\ {\small #3}
\end{tabular}
}

\newcommand{\R}{{\mathbb R}}
\newcommand{\N}{{\mathbb N}}

\newcommand{\abs}[1]{\lvert #1 \rvert}

\newcommand{\seq}{\subseteq}
\newcommand{\eps}{\varepsilon}
\newcommand{\epshat}{\hat{\eps}}
\newcommand{\Alg}{\mathcal{A}}
\renewcommand{\epsilon}{\eps}
\newcommand{\defeq}{\coloneqq}

\DeclareMathOperator{\E}{\mathbb{E}}
\newcommand{\1}{\mathbf{1}}

\newcommand{\Bits}{\{0,1\}}

\newcommand{\Fits}{\{-1,1\}}

\newcommand{\pE}{\tilde{\E}}
\newcommand{\cH}{\mathcal H}

\newcommand{\poly}{\operatorname{poly}}
\newcommand{\val}{\operatorname{val}}

\newcommand{\mper}{\,.}
\newcommand{\mcom}{\,,}

\newcommand{\polylog}{\operatorname{polylog}}

\newcommand{\cQ}{\mathcal Q}

\newcommand{\cL}{\mathcal{L}}
\newcommand{\Bias}{\operatorname{Bias}}

\newcommand{\sgn}{\text{sgn}}
\newcommand{\LPN}{\mathsf{XOR}}

\newcommand{\SoS}{\mathsf{SoS}}

\newcommand{\predict}{\operatorname{Predict}}
\newcommand{\pred}{\operatorname{pred}}
\newcommand{\prune}{\operatorname{Prune}}
\usepackage{amssymb} 

\definecolor{citeblue}{HTML}{0055cc}
\hypersetup{
  urlcolor=citeblue,
  linkcolor=NavyBlue,
  citecolor=PineGreen,
}

\renewcommand{\tilde}{\widetilde}

\renewcommand{\emptyset}{\varnothing}

\newcommand{\literalneg}{b}
\newcommand{\NP}{\mathsf{NP}}
\renewcommand{\P}{\mathsf{P}}

\begin{document}

\title{Sum-of-Squares Refutation Implies Planted Recovery for Semirandom CSPs}
\author{
\begin{tabular}[h!]{ccc}
     \FormatAuthor{Arpon Basu}{arpon.basu@princeton.edu}{Princeton University} &
     \FormatAuthor{Andrew D.\ Lin}{andrewlin@princeton.edu}{Princeton University}& \FormatAuthor{Peter Manohar\thanks{Supported by NSF Grant DMS-2424441.}}{pmanohar@ias.edu}{The Institute for Advanced Study} 
\end{tabular}
}%
\date{\today}


\maketitle

\vspace{-0.3cm}
\begin{abstract}
We give a $n^{O(\ell)}$-time algorithm for solving semirandom planted instances of a $k$-ary Boolean constraint satisfaction problem (CSP) with $m \gtrsim \left(\frac{n}{\ell}\right)^{k/2} \ell \polylog(n)$ constraints, for any choice of the runtime parameter $\ell$. Up to $\operatorname{polylog}(n)$ factors, this matches the known runtime vs.\ number of constraints trade-off for refuting semirandom CSPs, as well as the trade-off for solving fully random planted CSPs.

Prior work on designing algorithms for semirandom CSPs follows a high-level blueprint of decomposing the constraint hypergraph $\cH$ of the instance into pieces, each with an explicit pseudorandom property that allows the standard algorithm for random CSPs to succeed on every piece. We depart considerably from this framework and instead implicitly \emph{learn} such a decomposition, without appealing directly to explicit properties such as expansion.

More generally, we construct such a learning algorithm from the assumption that semirandom $k$-XOR has Sum-of-Squares certificates of refutation. This yields the broader implication that any Sum-of-Squares refutation algorithm for semirandom $k$-XOR gives a corresponding algorithm for solving semirandom planted CSPs, and instantiating this reduction with known Sum-of-Squares algorithms for refuting semirandom $k$-XOR gives our main algorithm. To the best of our knowledge, this is the first reduction from search to refutation for (semi)random CSPs.
\end{abstract}

\thispagestyle{empty}
\clearpage
 \microtypesetup{protrusion=false}
  \tableofcontents{}
  \microtypesetup{protrusion=true}

\thispagestyle{empty}
\clearpage

\pagestyle{plain}
\setcounter{page}{1}
\section{Introduction}
\label{sec:intro}
Understanding the computational complexity of constraint satisfaction problems (CSPs) is a fundamental goal in theoretical computer science. 
For worst-case CSPs, a long line of work has led to a rich and beautiful theory that gives us a deep understanding of their computational difficulty.
 To give a brief summary, deciding any CSP is either in $\P$ or is $\NP$-complete~\cite{Cook71,Levin73,Schaefer78,Bulatov17,Zhuk20}. For any $\NP$-complete CSP, the PCP theorem~\cite{AroraS98,AroraLMSS98,Dinur07} implies that it is also hard to approximate. Stronger variants of the PCP theorem~\cite{Hastad01} imply strong hardness-of-approximation results. For example, it is $\NP$-hard to decide if a $k$-$\mathsf{XOR}$ instance ($k\geq 3$) has an assignment satisfying $\geq 99\%$ of the equations, or if every assignment satisfies $\leq 51\%$ of the equations. Furthermore, $\NP$-hardness persists even when the number of equations is as large as $O(n^{k-1})$~\cite{FotakisLP16}, where $n$ is the number of variables. On the other hand, ``maximally dense'' $k$-CSPs with $m \geq \Omega(n^k)$ constraints have a polynomial-time approximation algorithm~\cite{AroraKK95}.
 
 Due to the above difficulty of designing algorithms for CSPs in the worst case, and also in part due to reasons in cryptography~\cite{Goldreich00}, there has been considerable effort to design algorithms for CSPs in the average case, that is, random CSPs. The study of algorithms for random CSPs can be divided into two lines of work, \textbf{refutation}~\cite{GoerdtL03,Coja-OghlanGL07,AllenOW15,BarakM16,RaghavendraRS17,AbascalGK21,GuruswamiKM22,HsiehKM23} and \textbf{search}~\cite{BarthelHL02,JiaMS07,BogdanovQ09,CojaCF10,FeldmanPV15,GuruswamiHKM23,BasuHLM26,ChanOX26}, depending on whether the distribution over instances for the $k$-CSP is supported primarily on unsatisfiable instances or satisfiable ones. In the refutation problem, the algorithmic task is to output a certificate of unsatisfiability, whereas in the search problem, the algorithmic task is to find a satisfying assignment. While the search task may appear to be the more natural algorithmic one, it is easier to define the refutation problem because it has the more natural distribution over inputs: a uniformly random $k$-CSP with $m \geq \tilde{O}(n)$ constraints is unsatisfiable with high probability. To define the search problem, one must specify a distribution over $k$-CSPs that simultaneously has a planted satisfying assignment $x^*$ but still feels similar to a uniformly random CSP. Doing this requires some care to ensure that recovering the planted assignment is not a trivial algorithmic task.\footnote{For example, for some CSPs such as $k$-$\mathsf{SAT}$, if every clause is sampled uniformly randomly conditioned on being satisfied by $x^*$, then a simple ``literal counting'' argument suffices to recover $x^*$ when the instance only has $O(n \log n)$ constraints. In such cases, clauses have to be sampled non-uniformly from the set of clauses satisfying $x^*$ to pose an interesting algorithmic question. See \cref{def:semirandomplantedcsp} for details.}
 
However, in sharp contrast to the $\NP$-hardness of worst-case $k$-CSPs (even up to $O(n^{k-1})$ clauses), a line of work~\cite{GoerdtL03,Coja-OghlanGL07,AllenOW15,BarakM16} shows that one can refute uniformly random $k$-CSPs with $m \gtrsim n^{k/2}$ clauses in polynomial time. Furthermore,~\cite{RaghavendraRS17} shows that one can refute random $k$-CSPs with $m \gtrsim (n/\ell)^{k/2} \cdot \ell$ clauses in $n^{O(\ell)}$ time for any choice of the ``runtime parameter'' $1\leq\ell\leq n$, thus giving a smooth trade-off between runtime and number of clauses, as the runtime varies from polynomial to fully exponential when the number of clauses decreases from $n^{k/2}$ to $n$. This trade-off is known to be optimal for Sum-of-Squares algorithms~\cite{KothariMOW17}, and is also the basis of some cryptographic hardness assumptions~\cite{ApplebaumBW10}. 
More recently,~\cite{AbascalGK21,GuruswamiKM22,HsiehKM23} show that one can also refute \emph{semirandom} $k$-CSPs at the same runtime vs.\ number of constraints trade-off as~\cite{RaghavendraRS17}: a semirandom $k$-CSP is a CSP that is chosen with a worst-case constraint hypergraph (or scopes), while the literal negations are chosen independently and uniformly at random. That is, a semirandom $k$-CSP is a random $k$-CSP with an arbitrary constraint hypergraph.

A parallel line of results also holds for solving random planted CSPs. A long line of works~\cite{BarthelHL02,JiaMS07,BogdanovQ09,CojaCF10,FeldmanPV15}, culminating in~\cite{FeldmanPV15}, gives a polynomial-time algorithm to solve random planted $k$-CSPs with $m \gtrsim n^{k/2}$ clauses, and this was extended very recently by~\cite{BasuHLM26} to obtain the ``\cite{RaghavendraRS17}''-style trade-off between runtime and number of clauses. That is,~\cite{BasuHLM26}  gives an algorithm to solve random planted CSPs in $n^{O(\ell)}$ time with $m \gtrsim (n/\ell)^{k/2} \ell$ constraints. Thus, for random CSPs, the algorithms for refutation and search have essentially\footnote{Technically, this holds up to $\polylog(n)$ factors, which we have suppressed.} identical runtime vs.\ number of clauses trade-offs, even though they have been obtained in parallel lines of work and we have no formal reductions between the two.

Surprisingly, progress on designing algorithms for semirandom planted CSPs has been much more limited. Analogous to the definition of semirandom CSPs in the refutation case, a semirandom planted CSP is sampled similarly to a random planted CSP, only the constraint hypergraph is now worst-case. The only known algorithmic result is due to~\cite{GuruswamiHKM23}, which gives a $\poly(n)$-time algorithm for solving\footnote{Technically, the algorithm outputs an assignment satisfying a $(1-o(1))$-fraction of the equations, and this almost certainly cannot be avoided.} semirandom planted CSPs with $m \gtrsim n^{k/2}$ constraints. Despite several recent attempts~\cite{BasuHLM26,Kothari26}, we still do not have an algorithm that achieves a ``\cite{RaghavendraRS17}''-style runtime vs.\ constraints trade-off for semirandom planted CSPs.

The lack of a ``\cite{RaghavendraRS17}''-style trade-off for semi-random planted CSPs is aesthetically displeasing in its own right; we want to obtain the same complete picture of search algorithms for planted CSPs in the semirandom case like we do for random CSPs. However, more importantly, it exposes a more fundamental gap in our understanding of the average-case complexity of CSPs: \emph{we do not know how to relate the computational tractability/difficulty between refutation and search!} This gap in our understanding is made even more perplexing by the fact that the runtime vs.\ constraints trade-offs are nearly identical for both problems in the random case, even though known algorithms for search generally require significantly more technical ideas than their refutation counterparts. While there has been a recent attempt at unifying these two lines of work~\cite{AndersonMX26}, there is still no reduction known between refutation and search.

\parhead{Our contributions.} As the main contribution of this paper, we close this gap in our understanding by showing a \emph{generic reduction} that gives an algorithm for solving semirandom planted CSPs from Sum-of-Squares refutation certificates for semirandom $k$-XOR.
Instantiating this reduction with known ``\cite{RaghavendraRS17}''-style Sum-of-Squares algorithms for refuting semirandom $k$-XOR yields an ``\cite{RaghavendraRS17}''-style algorithm for solving semirandom planted CSPs. 

We note that strictly speaking, this is not a reduction from, say, the semirandom planted $k$-$\mathsf{SAT}$ problem to Sum-of-Squares refutation for semirandom $k$-$\mathsf{SAT}$, but rather $k$-XOR. However, we note that \emph{all} Sum-of-Squares refutation algorithms for general Boolean CSPs go via a reduction from a general $k$-CSP to semirandom $k$-XOR, and similarly algorithms for planted CSPs~\cite{FeldmanPV15,GuruswamiHKM23,BasuHLM26}, including the algorithm in this paper, also go via a reduction from the general case to $k$-XOR. Thus, the asymmetry between the predicate of the $k$-CSP of the planted problem versus the $k$-XOR predicate for refutation should be viewed only a mild technicality.

\subsection{Our results}
We now turn to giving more formal statements of our results. We will first explain our results for $k$-XOR before we give our results for general  Boolean $k$-CSPs. Before we continue, we recall the definition of semirandom noisy planted $k$-XOR, as defined in~\cite{GuruswamiHKM23}.

\begin{definition}[$k$-XOR instance]
\label{def:kxor}
An instance $\Psi$ of $k$-XOR with $n$ variables and $m$ constraints is defined by the data $(\cH, \{b_C\}_{C \in \cH})$, where (1) $\cH \subseteq {[n] \choose k}$ is an arbitrary multiset, and (2) $b_C \in \Fits$ for each $C \in \cH$. This data represents the system of constraints $\{x_C = b_C\}_{C \in \cH}$, where $x_C \defeq \prod_{i \in C} x_i$.

We let $\Psi(x) \defeq \frac{1}{2} + \frac{1}{2m} \sum_{C \in \cH} b_C x_C$ denote the polynomial where for every $x \in \Fits^n$, $\Psi(x)$ is the fraction of constraints satisfied by $x$. We call this polynomial the \emph{instance polynomial} of $\Psi$.
\end{definition}
\begin{definition}[Semirandom planted noisy $k$-XOR]
\label{def:semirandomkxor}
Let $\cH \subseteq {[n] \choose k}$ be an arbitrary multiset, and let $x^* \in \Fits^n$. Let $\LPN(\cH, x^*, \eps)$ denote the distribution on $k$-XOR instances obtained by, for each $C \in \cH$, sampling bits $r_C \sim \Bias(\eps)$ independently for each $C \in \cH$, and then adding the constraints $\{x_C = b_C\}_{C \in \cH}$, where $b_C = r_C x^*_C$. Here, $\Bias(\eps)$ is the distribution over $\Fits$ that is $1$ with probability $\frac{1 + \eps}{2}$, and otherwise is $-1$, and $x_C \defeq \prod_{i \in C} x_i$. A semirandom planted noisy $k$-XOR instance $\Psi$ is an instance drawn from $\LPN(\cH, x^*, \eps)$, for some choice of $\cH$, $x^*$, and $\eps > 0$. We note that $\Psi$ is represented by the data $\Psi = (\cH, \{b_C\}_{C \in \cH})$, where the $b_C$'s are the right-hand sides of the constraints.
\end{definition}

Our main result shows that if degree-$\ell$ Sum-of-Squares is capable of certifying that $\val(\Phi) \leq \frac{1}{2} + o(1)$ when $\Phi \sim \LPN(\cH, \1^n, 0)$, i.e., the right-hand sides $b_C$ of the instance $\Phi$ are chosen uniformly at random from $\Fits$, then there is an $n^{O(\ell)}$-time algorithm that, when given an instance $\Psi \sim \LPN(\cH, x^*, \eps)$, is essentially capable of \emph{de-noising} the constraints, provided that we approximately know $\eps$.
\begin{mtheorem}[Informal \cref{thm:mainformal}]
\label{mthm:main}
Let $k, n \in \N$. There is an algorithm $\Alg$ that takes as input a ``runtime parameter'' $k/2 \leq \ell \leq n$, a ``bias guess'' $\epshat \in (0,1)$, and a $k$-XOR instance $\Psi = (\cH, \{b_C\}_{C \in \cH})$ with constraint hypergraph $\cH$ of size $\abs{\cH} = m \leq n^{O(k)}$, and in time $n^{O(\ell)}$ outputs a partition $\cH_{+} \sqcup \cH_{-} \sqcup \cH_{0}$ of $\cH$ with the following guarantee. Suppose that, with probability $1 - o(1)$ over $\Phi \sim \LPN(\cH, \1^n, 0)$, degree-$\ell$ Sum-of-Squares can certify that $\val(\Phi) \leq \frac{1}{2} + o(1/\log(mn))$. Then, with probability $1 -  O(m\abs{\epshat - \eps}) - o(1)$, it holds that (1) for every $C \in \cH_{+}$, we have $b_C = x_C$, (2) for every $C \in \cH_{-}$, we have $b_C = -x_C$, and (3) $\cH_{0}$ is a set of size $\leq o(m)$.
\end{mtheorem} 
Applying known results on semirandom $k$-XOR refutation using Sum-of-Squares~\cite{AbascalGK21,GuruswamiKM22,HsiehKM23}, we obtain the following corollary.
\begin{corollary}
\label{cor:main}
Let $k, n \in \N$. There is an algorithm $\Alg$ that takes as input a ``runtime parameter'' $k/2 \leq \ell \leq n$, a ``bias guess'' $\epshat \in (0,1)$, and a $k$-XOR instance $\Psi = (\cH, \{b_C\}_{C \in \cH})$ with constraint hypergraph $\cH$ of size $\abs{\cH} = m \leq n^{O(k)}$, and in time $n^{O(\ell)}$ outputs a partition $\cH_{+} \sqcup \cH_{-} \sqcup \cH_{0}$ of $\cH$ with the following guarantee. Suppose that the input $\Psi$ is drawn from $\LPN(\cH, x^*, \eps)$ and $\abs{\cH} = m \geq \left(\frac{n}{\ell}\right)^{k/2} \ell \polylog(n)/\eps^4$. Then, with probability $1 -  O(m\abs{\epshat - \eps}) - o(1)$, it holds that (1) for every $C \in \cH_{+}$, we have $b_C = x_C$, (2) for every $C \in \cH_{-}$, we have $b_C = -x_C$, and (3) $\cH_{0}$ is a set of size $\leq o(m)$.
\end{corollary}
The output sets $\cH_{+}$, $\cH_{-}$ of \cref{mthm:main,cor:main} should be interpreted as the algorithm's guesses for constraints where $r_C = 1$ and $r_C = -1$, respectively, with $\cH_{0}$ denoting a set of constraints where the algorithm is allowed to ``give up'' with no penalty, albeit subject to the requirement that it cannot ``give up'' on too many constraints. The strong guarantee achieved by \cref{mthm:main,cor:main} is that with high probability, the algorithm's implicit guess is \emph{correct} for every $C \in \cH_{+} \sqcup \cH_{-}$ \emph{simultaneously}, and $\cH_{+} \sqcup \cH_{-}$ is the bulk of the constraints. This means that by changing the sides of the right-hand sides appropriately and running Gaussian elimination on the constraints in $\cH_{+} \sqcup \cH_{-}$, we can easily obtain an assignment $x \in \Fits^n$ where $\prod_{i \in C} x_i = \prod_{i \in C} x^*_i$ for all $C \in \cH_{+} \sqcup \cH_{-}$, and since this is the bulk of the constraints, we then have $\Psi(x) \geq \Psi(x^*) - o(1)$.

Notice that if we are given the hidden ``noise bits'' $\{r_C\}_{C \in \cH}$ (as defined in \cref{def:semirandomkxor}) for the instance $\Psi \sim \LPN(\cH, x^*, \eps)$, then we can trivially recover $x^*_C$ for every $C \in \cH$, which is all the information that the hypergraph $\cH$ can yield about $x^*$. Because $\cH_{0}$ is ``small'', the algorithm recovers $x^*_C$ for almost all $C \in \cH$, which is intuitively almost all of the information encoded about $x^*$ in the instance $\Psi$. As pointed out in~\cite{GuruswamiHKM23}, there are simple examples that show that we cannot hope to recover $x^*_C$ for all $C \in \cH$, allowing the algorithm to discard some small subset of $\cH$ is necessary.

However, there is one caveat about the algorithm in \cref{mthm:main}: it requires a very good estimate $\epshat$ of $\eps$ in order for the algorithm to succeed with high probability. This is not an issue if we are only interested in recovering a good assignment $x$ to the instance $\Psi$, as we can try all values of $\epshat = t/m^2$ for $t = 1, \dots, m^2$, and one of them is guaranteed to be $O(1/m^2)$-close to $\eps$. The following corollary captures this algorithm, which successfully recovers a near-optimal assignment to a semirandom planted noisy $k$-XOR instance.
\begin{corollary}[Informal]
\label{cor:semirandomkxor}
Let $k, n \in \N$. There is an algorithm $\Alg$ that takes as input a ``runtime parameter'' $k/2 \leq \ell \leq n$ and a $k$-XOR instance $\Psi = (\cH, \{b_C\}_{C \in \cH})$ with constraint hypergraph $\cH$ of size $\abs{\cH} = m \leq n^{O(k)}$, and in time $n^{O(\ell)}$ outputs an assignment $x \in \Fits^n$ with the following guarantee. Suppose that the input $\Psi$ is drawn from $\LPN(\cH, x^*, \eps)$ and $\abs{\cH} = m \geq \left(\frac{n}{\ell}\right)^{k/2} \ell \poly(\log(n), \frac{1}{\eps})$. Then, with probability $1 - o(1)$, it holds that $\Psi(x) \geq \Psi(x^*) - o(1)$.

Note that with probability $1 - o(1)$, we have $\Psi(x^*) \geq \frac12+\frac{\eps}{2} - o(1)$ and $\val(\Psi) \leq \frac12+\frac{\eps}{2} + o(1)$, so $\Psi(x) \geq \val(\Psi) - o(1)$ with probability $1 - o(1)$.
\end{corollary}
We prove \cref{mthm:main,cor:main,cor:semirandomkxor} in \cref{sec:kxor}.

\parhead{Extension to general Boolean CSPs.} As shown by the reduction in~\cite[Section 4]{GuruswamiHKM23},\footnote{Technically, the reduction requires a small modification due to the presence of the ``bias guess'' $\epshat$ in \cref{mthm:main}. We explain the slight modification in \cref{sec:generalcsp}.} any algorithm for semirandom planted noisy $k$-XOR with the strong output guarantees as in \cref{mthm:main} can be extended to give an algorithm for solving any semirandom planted Boolean $k$-CSP. We thus obtain analogous algorithms for general semirandom planted CSPs. Before we can state the guarantees of our algorithm, we recall the definitions of semirandom planted CSPs, as defined in~\cite{GuruswamiHKM23}.

\begin{restatable}[$k$-ary Boolean CSP]{definition}{defbooleancsp}
A CSP instance $\Psi$ with a $k$-ary predicate $P \colon \Fits^k \to \Bits$ is a set of $m$ constraints on variables $x_1,\dots,x_n$ of the form $P(\literalneg_{C,1} x_{i_1}, \literalneg_{C,k} x_{i_2}, \ldots, \literalneg_{C,k} x_{i_k}) = 1$, where $C = (i_1, \ldots, i_k) \in [n]^k$ ranges over a collection $\cH$ of \emph{scopes} (or clause structure) of $k$-tuples of $n$ variables and $\literalneg_C \in \Fits^k$ are ``literal negations'', one for each $C$ in $\cH$. We additionally allow $\cH$ to be a multiset, i.e., that multiple clauses can contain the same ordered tuple of variables.

We let $\Psi(x)$ denote the fraction of constraints satisfied by an assignment $x \in \Fits^n$, and we define the \emph{value} of $\Psi$, $\val(\Psi)$, to be $\max_{x \in \Fits^n} \Psi(x)$.
\end{restatable}

\begin{definition}[Instance polynomial]
\label{def:instancepoly}
Given a $k$-CSP instance $\Psi$ with predicate $P$, we define the instance polynomial $\Psi$ as follows
\begin{flalign*}
\Psi(x) \defeq \frac{1}{m} \sum_{C = (i_1, \dots, i_k) \in \cH} P(b_{C,1} x_{i_1}, \dots, b_{C,2} x_{i_k}) \mper
\end{flalign*}
Note that $\Psi$ is a degree $\leq k$ polynomial, and that $\Psi(x)$ is the fraction of constraints satisfied by an assignment $x$.
\end{definition}

A semirandom planted CSP is defined as follows.
\begin{definition}[Semirandom planted $k$-ary Boolean CSPs]
\label{def:semirandomplantedcsp}
Let $P \colon \Fits^k \to \Bits$ be a predicate. We say that a distribution $\cQ$ over $\Fits^k$ is a \emph{planting distribution for $P$} if $\Pr_{y \sim \cQ}[P(y) = 1] = 1$.

We say that an instance $\Psi$ with predicate $P$ is a \emph{semirandom planted instance} with \emph{planting distribution} $\cQ$ if it is sampled from a distribution $\Psi(\cH, x^*, P, \cQ)$ where
\begin{enumerate}[(1)]
    \item The planted assignment $x^* \in \Fits^n$ is arbitrary;
    \item the set $\cH \subseteq [n]^k$ is an arbitrary multiset. We let $m = \abs{\cH}$;
    \item for each $C = (i_1, \dots, i_k)\in \cH$, the literal negations $\literalneg_C$ are sampled by $\literalneg_C \sim \cQ(\literalneg(C) \odot (x^*_{i_1}, \dots, x^*_{i_k}))$, where ``$\odot$'' denotes the element-wise product of two vectors.
    That is, $\Pr[\literalneg_C = y] = \cQ(y \odot (x^*_{i_1}, \dots, x^*_{i_k}))$ for each $y \in \Fits^k$.
    Then, add the constraint 
    \[P(\literalneg_{C,1} x_{i_1}, \literalneg_{C,2} x_{i_2}, \ldots, \literalneg_{C,k} x_{i_k}) = 1\] 
    to the instance $\Psi$.
\end{enumerate}
Because $\cQ$ is supported only on satisfying assignments to $P$, it follows that if $\Psi \sim \Psi(\cH, x^*, P, \cQ)$, then $\Psi(x^*) = 1$ with probability $1$, i.e., $x^*$ satisfies $\Psi$.
\end{definition}

\newcommand{\phisp}{\psi^{(S,+)}}
\newcommand{\phism}{\psi^{(S,-)}}

\begin{mtheorem}[Informal]
\label{mthm:main2}
Let $k, n \in \N$. There is an algorithm $\Alg$ that takes as input a ``runtime parameter'' $k/2 \leq \ell \leq n$ and a $k$-CSP $\Psi$ with predicate $P$ and constraint hypergraph $\cH$ of size $\abs{\cH} = m \leq n^{O(k)}$, and in time $n^{O(\ell)}$ outputs an assignment $x \in \Fits^n$ with the following guarantee. Suppose that, with probability $1 - o(1)$ over $\Phi \sim \LPN(\cH, \1^n, 0)$, degree-$\ell$ Sum-of-Squares can certify that $\val(\Phi) \leq \frac{1}{2} + o(1)$. Then, if $\Psi \sim \Psi(\cH, x^*, P, \cQ)$ for any $x^*$, with probability $1 -  o(1)$, it holds that $\Psi(x) \geq 1 - o(1)$.
\end{mtheorem}
Applying known results on semirandom $k$-XOR refutation using Sum-of-Squares, gives the corollary:
\begin{corollary}
\label{cor:semirandomcsp}
Let $k, n \in \N$. There is an algorithm $\Alg$ that takes as input a ``runtime parameter'' $k/2 \leq \ell \leq n$ and a $k$-CSP $\Psi$ with predicate $P$ and constraint hypergraph $\cH$ of size $\abs{\cH} = m \leq n^{O(k)}$, and in time $n^{O(\ell)}$ outputs an assignment $x \in \Fits^n$ with the following guarantee. Suppose that the input $\Psi$ is drawn from $\Psi(\cH, x^*, P, \cQ)$ and $\abs{\cH} = m \geq \left(\frac{n}{\ell}\right)^{k/2} \ell \polylog(n)$. Then, with probability $1 - o(1)$, it holds that $\Psi(x) \geq 1 - o(1)$.
\end{corollary}
We prove \cref{mthm:main2,cor:semirandomcsp} in \cref{sec:generalcsp}.

\subsection{Technical Overview}
\label{sec:overview}
In this subsection, we give an overview of the proof of \cref{mthm:main}. Let $\Psi = (\cH, \{b_C\}_{C \in \cH})$ be the input to the algorithm.

\parhead{The barriers of current approaches.} The  strategy employed by~\cite{GuruswamiHKM23} is to partition $\cH$ into sets $\cH_0 \sqcup \cH_1 \sqcup \dots \sqcup \cH_t$, where $\cH_0$ is the set of bad hyperedges that will be outputted in \cref{mthm:main}, and for every constraint set $\cH_i$ in the partition, there is an algorithm that recovers $x^*_C$ for all $C \in \cH_i$ correctly and simultaneously with high probability when given $\Psi_i \sim \LPN(\cH_i, x^*, \eps)$. As the subinstance of $\Psi$ with hyperedges $\cH_i$ is clearly distributed as $\LPN(\cH_i, x^*, \eps)$, the result follows.

The way this strategy is executed in~\cite{GuruswamiHKM23} is by using \emph{expander decomposition}. For simplicity, let us assume that $k$ is even, as the proof of~\cite{GuruswamiHKM23} is substantially simpler when $k$ is even. They first convert the hypergraph $\cH$ to a graph using the basic, or level $k/2$, Kikuchi graph, which is a graph $G_{\cH,k/2}$ with vertices $V = {[n] \choose k/2}$ and an edge $(S,T)$ if $S \cup T = C \in \cH$. By using expander decomposition, they decompose $G_{\cH, k/2}$ into expanders with spectral gap $1/\log \abs{V} \approx 1 /\log n$, and they argue that the basic SDP for $2$-XOR successfully ``de-noises'' all edges if the graph is an expander.

The issue with this strategy is that it seems difficult to generalize it to the general case of $n^{O(\ell)}$-time algorithms. The natural approach here is to try to use the level $\ell$ Kikuchi graph, which is a graph $G_{\cH, \ell}$ with vertices $V = {[n] \choose \ell}$ and edges $(S,T)$ if $S \oplus T = C \in \cH$, where $S \oplus T$ is the symmetric difference of $S$ and $T$. However, the expander decomposition strategy no longer works, as the spectral gap that one can ensure is only $1/\log N \sim 1/\ell \log n$. This ends up losing an extra $\poly(\ell)$ factor in the number of constraints, and which does not yield the $n^{O(\ell)}$ vs.\ $m \sim (n/\ell)^{k/2} \ell$  constraints trade-off that we are targeting.

More recently, there have been two attempts, first by~\cite{BasuHLM26} and then by~\cite{Kothari26} to extend the strategy of~\cite{GuruswamiHKM23} to give an algorithm with an ``\cite{RaghavendraRS17}''-style trade-off. However, both attempts have only succeeded in the case of random $k$-CSPs, and the latter work of~\cite{Kothari26} even has the additional restriction that $k$ must be even, which is typically the case that is substantially easier to prove.

The algorithm in~\cite{BasuHLM26} is a $2$-step algorithm that first finds an assignment $\hat{x}$ that is close to $x^*$ while holding out $O(n \log n)$ of the random constraints, and then in the second step it rounds $\hat{x}$ to exactly recover $x^*$ using these extra constraints. The key idea in the first step is to \emph{not} use the Kikuchi graph $G_{\cH,\ell}$, but instead study a direct notion of expansion on $\cH$. This notion, called ``$(\ell,\lambda)$-certifiable expansion'' in~\cite{AndersonMX26}, says that
\begin{equation*}
\left|\pE\left[\E_{C\sim\cH}[x_C] - \left(\frac{\langle x, \1\rangle}{n}\right)^k\right]\right|\leq \lambda \mcom
\end{equation*}
for every degree-$\ell$ pseudo-expectation $\pE$. Certifiable expansion is a natural generalization of spectral expansion, which is for homogeneous degree $2$ polynomials, to homogeneous degree $k$ polynomials. It is therefore natural to wonder if one can replace the spectral expansion criteria in~\cite{GuruswamiHKM23} with certifiable expansion. It turns out, however, that this is too much to hope for. The work of~\cite{ChanOX26,AndersonMX26} shows that if a $k$-CSP $\Psi$ has a constraint hypergraph $\cH$ that is $(\ell, o(1))$-certifiably expanding, then there is an $n^{O(\ell)}$-time algorithm that finds a near-optimal assignment $x$, i.e., with $\Psi(x) \geq \val(\Psi) - o(1)$, and this algorithm succeeds even if the literal negations of $\Psi$ are worst case, and may be chosen depending on the hypergraph $\cH$. In other words, if we could decompose any hypergraph $\cH$ into certifiably expanding pieces while only discarding a small fraction of hyperedges, then we would be able to find a near-optimal assignment to any worst-case CSP, which would imply that $\P = \NP$. Thus, certifiable expansion is too strong to be used for semirandom planted CSPs.

The work of~\cite{Kothari26} reproves the main result of~\cite{BasuHLM26} in the restricted case of even $k$ with a different approach: it shows that if $\cH$ is random, then $G_{\cH,\ell}$ is a spectral sparsifier of the Johnson scheme. The hope expressed in~\cite{Kothari26} is that one could decompose an arbitrary hypergraph $\cH$ into components that are each approximately Johnson, and then the analysis of~\cite{GuruswamiHKM23} would then immediately give an algorithm for semirandom planted CSPs. However, this hope is too good to be true, for the same reason as certifiable expansion. Indeed, it is not too difficult to observe that if $G_{\cH,\ell}$ is a spectral sparsifier of the Johnson scheme with $o(1)$ relative error, then $\cH$ is $(\ell, o(1))$-certifiably expanding.
Thus, the arguments of~\cite{ChanOX26,AndersonMX26} again imply that this is too strong of a condition to be used in the semirandom case.

In essence, both certifiable expansion and Johnson spectral sparsification are properties that rely very heavily on having a random $\cH$, and are too strong to extend to the semirandom setting of a worst-case $\cH$.

\parhead{Our approach: learning the set of bad hyperedges.} The key idea behind our approach is very simple: rather than attempting to find the ``bad hyperedges'' $\cH_0$ and a decomposition of the remaining hyperedges into ``good pseudorandom pieces'' $\cH_i$, we will instead \emph{learn} the bad hyperedges $\cH_0$ directly. For simplicity, in the below exposition we will assume that the algorithm knows $\eps$ exactly. We note that in \cref{mthm:main}, this is not required, and the algorithm is only guaranteed to work well if it is given an estimate $\epshat$ that is close to $\eps$. 

To design the learning algorithm, we consider the following experiment. Suppose we sample $\Psi' = (\cH, \{b'_C\}_{C \in \cH}) \sim \LPN(\cH, \1^n, \eps)$, which we note is identically distributed as $\Psi$ (since we know $\eps$) except the planted assignment is $\1^n$ instead of $x^*$. Now, we can compute the degree-$\ell$ pseudo-expectation that maximizes $\E[\Psi'(x)]$. By appealing to known results for semirandom $k$-XOR refutation, we can show (see proof of \Cref{lem:predalg} in \Cref{sec:predalg})
that $\pE$ has $\E_{C \sim \cH}[\pE[x_C x^*_C]] \geq 1 - o(1)$. As $\pE[x_C x^*_C] \in [-1,1]$ by the Booleanity constraints, it follows that at least $(1 - o(1))m$ hyperedges have $\pE[x_C x^*_C] \geq 1 - o(1)$.

The above analysis shows that if we use $\sgn(\pE[x_C x^*_C])$ as a ``guess'' for whether or not $b'_C = x^*_C$, i.e., whether or not the hyperedge $C$ is noisy in $\Psi'$, then every time we run this experiment, there are at least $(1 - o(1))m$ hyperedges where this estimator is correct.
Note that this not sufficient to imply \cref{mthm:main} because the output of \cref{mthm:main} is not permitted to make any mistakes.  If we run the estimator directly on the input instance $\Psi$, it will have correct predictions for most hyperedges, but we do not know which hyperedges are wrong. (We cannot union bound over all $m$ hyperedges because the $o(1)$ term can only be as small as $1/\polylog(n)$.) That is, we need to determine the set $\cH_0$ of hyperedges where the estimator is untrustworthy.

Let $\cH'$ denote the set of ``good'' hyperedges where the estimator is correct with probability at least $2/3$ over the draw of $\Psi'$. Imagine that we are given $t$ independent samples $\Psi^{(1)}, \dots, \Psi^{(t)} \sim \LPN(\cH, x^*, \eps)$ from the same distribution as $\Psi$. (Note that we cannot actually draw these samples, as we do not know $x^*$.) The ``noise predictor'' immediately gives a prediction for each $x^*_C$ in each subsample, and we know that if $C \in \cH'$ is ``good'', then the $t$ predictions are i.i.d.\ and each correct with probability $\geq 2/3$. If $t = O(\log n)$, then taking the majority vote is correct with probability $1 - 1/\poly(n)$, and this is high enough to take a union bound over all hyperedges in $\cH'$.

There are still two components missing from the full proof. First, we need to argue that we can find the set $\cH'$ algorithmically. This is fairly straightforward, as by generating the fake samples $\Psi'$ (which we can do, because they have planted assignment $\1^n$) we can obtain an empirical estimate for the probability that estimator is correct on each hyperedge. By taking all hyperedges $C$ with empirical probability larger than some appropriate cut-off, we obtain a subset $\cH''$ of size at least $(1 - o(1))m$ and we also have $\cH'' \subseteq \cH'$ with high probability. This allows us to find a large subset of the good hyperedges, which is sufficient. We set $\cH_0 = \cH \setminus \cH''$.

The far more pressing issue, however, is that we cannot actually generate the samples $\Psi^{(1)}, \dots, \Psi^{(t)}$ ourselves because we do not know $x^*$. The key technical lemma, \cref{lem:splitting}, shows that we actually \emph{can} generate these samples from the single sample $\Psi$, at the cost of decreasing the bias parameter $\eps$. In more detail, \cref{lem:splitting} shows the following simple statement. Suppose we are given a bit $b \in \Fits$ generated as $b = s r$ where $s \in \Fits$ is secret and $r \sim \Bias(\eps)$. Then, there is an algorithm running in $2^{O(t)}$ time that outputs $(b^{(1)}, \dots, b^{(t)})$ that is jointly distributed as i.i.d.\ samples from $\Bias(\eps/2t)$. This allows us to split the single sample $\Psi$ from $\LPN(\cH, x^*, \eps)$ into samples $\Psi^{(1)}, \dots, \Psi^{(t)}$ from $\LPN(\cH, x^*, \eps/2t)$. Running the predictor algorithm with samples generated from $\LPN(\cH, \1^n,\eps/2t)$ then finishes the algorithm. We note that \cref{lem:splitting} is closely related to more general ``cloning theorems'' in literature, see, for e.g.\, \cite[Lemma~14]{BrennanBH19} and \cite[Lemma~22]{BrennanB20}. We nevertheless prove \cref{lem:splitting} to keep our presentation self-contained.

Finally, we note that while the algorithm does not know $\eps$ itself, one can show that if we run the algorithm using $\epshat$ to generate our samples, rather than the true $\eps$, then all of the above distributions using $\epshat$ are $O(\poly(m)\abs{\epshat - \eps})$-close in TV distance to the ``true target'' distributions that use $\eps$. This allows us to make do with a ``good enough'' estimate of $\eps$, which we can guess by trying all possibilities.

The full proof of \cref{mthm:main} and its related corollaries \cref{cor:main,cor:semirandomkxor} is presented in \cref{sec:kxor}. We prove the guarantees of the ``noise predictor'' (\cref{lem:predalg}) in \cref{sec:predalg}, and we prove the sample splitting lemma (\cref{lem:splitting}) in \cref{sec:splitting}. We prove \cref{mthm:main2}, the generalization to semirandom planted $k$-CSPs, and its related corollary \cref{cor:semirandomcsp} in \cref{sec:generalcsp}.
\section{Preliminaries}
We state some notation and background that we will use in the proofs of our results.

We define the distribution $\Bias(\eps)$, as follows.
\begin{definition}[Biased random bit]
    Let $\eps\geq 0$. Let $\Bias(\eps)$ to be the distribution that is equal to $1$ with probability $(1+\eps)/2$ and $-1$ otherwise.
\end{definition}
\subsection{Sum-of-Squares background}
\label{sec:sos}

We recall some basic facts about SoS (see \cite{BarakS16,FlemingKP19} for further details). Define $\R[x_1, \ldots, x_n]_{\leq t}$ to be the set of polynomials in $\R[x_1, \ldots, x_n]$ of degree $\leq t$.
\begin{definition}[Pseudo-expectations over the hypercube]
\label{def:sosaxioms}
    For any $d\geq 2$, a degree $d$ pseudo-expectation $\pE$ over $\Fits^n$ is a linear functional $\pE: \R[x_1, \ldots, x_n]_{\leq d}\to\R$ satisfying the following properties:
    \begin{enumerate}
        \item (Normalization) $\pE[1] = 1$, 
        \item (Booleanity) $\pE[fx_i^2] = \pE[f]$ for all $i\in[n], f\in\R[x_1, \ldots, x_n]_{\leq d - 2}$, 
        \item (Positivity) $\pE[f^2]\geq 0$ for all $f\in\R[x_1, \ldots, x_n]_{\leq d/2}$.
    \end{enumerate}
    Finally, denote by $\SoS_d$ $(\SoS_{\geq d})$ the set of all degree $d$ ($\geq d$) pseudo-expectations over $\Fits^n$.
\end{definition}
We will make a slight abuse of terminology and use the phrase ``pseudo-expectation'' to mean a pseudo-expectation over $\Fits^n$. Note that for any degree-$t$ multlinear polynomial $f \colon \Fits^n \to \R$, it holds that $\max_{x \in \Fits^n} f(x) \leq \sup_{\pE\in\SoS_d}\pE[f]$ for all $t \leq d\leq n$.

Given a degree-$t$ polynomial $f$, the SoS algorithm can compute all moments up to degree $\leq d$ of some pseudo-expectation in $\sup_{\pE\in\SoS_d}\pE[f]$ in $n^{O(d)}$ time.
\begin{fact}[SoS Algorithm (Corollary 3.40 in \cite{FlemingKP19})]
\label{sosfkp}
    Let $f = f(x_1, \ldots, x_n)$ be a polynomial of degree $t$ with rational coefficients such that each coefficient has $\poly(n)$ bit complexity. Then for any $d\geq t$, there exists an algorithm that, on input $f$ and $d$, runs in time $n^{O(d)}$ and outputs $\{\alpha_S\}_{S\in\binom{[n]}{\leq d}}$, where $|\alpha_S - \pE_\mu[x_S]|\leq 4^{-n}$ for all $S\in\binom{[n]}{\leq d}$, and $\pE_\mu \in \arg\max_{\pE\in\SoS_d}\pE[f]$. In particular in $n^{O(d)}$ time one can compute $\alpha:= \pE_\mu[f]$ such that $\alpha$ satisfies $\alpha \geq \beta - 2^{-n}$, where $\beta:= \max_{x\in\Fits^n}f(x)$. 
\end{fact}

We will use the fact that Sum-of-Squares is capable of certifying that a semirandom $k$-XOR instance has value $\leq 1/2 + o(1)$, which is shown in~\cite{AbascalGK21,GuruswamiKM22,HsiehKM23}.
\begin{fact}[Theorem~4.1 of \cite{HsiehKM23}]
\label{fact:hkmrefutation}
    Fix $2\leq k\in \mathbb{N}$, and let $2k \leq \ell \leq n/8$. Let $\cH$ be a $k$-uniform hypergraph with $n$ vertices and $m$ hyperedges, and let $\Psi \sim (\cH, \1^n, 0)$ be a semirandom $k$-XOR instance, where the right-hand sides of each constraint are independently chosen from $\Fits$. If $m \geq O_A(1)^k \cdot(\frac{n}{\ell})^{\frac{k}{2}} \ell \log n \cdot \delta^{-4}$ for $\delta \in (0,1/2)$ and $A>0$, then with probability $1-n^{-A}$ over $\Psi$, it holds that $\pE[\Psi(x)] \leq \frac{1}{2} + \delta$ for every degree-$2\ell$ pseudo-expectation $\pE$ (\cref{def:sosaxioms}).
\end{fact}

\newcommand{\heps}{\hat{\eps}}
\newcommand{\xii}{\xi^{(i)}}
\newcommand{\xiii}{\xi^{(i),2}}
\newcommand{\zi}{\zeta^{(i)}}
\newcommand{\zii}{\zeta^{(i),2}}
\newcommand{\predi}{\pred^{(i)}}
\newcommand{\pei}{\pE^{(i)}}
\section{Proof of \cref{mthm:main}}
\label{sec:kxor}
In this section, we prove the formal versions of \cref{mthm:main,cor:main,cor:semirandomkxor}. We will first prove \cref{thm:mainformal} stated below, which is the formal version of \cref{mthm:main}.
\begin{theorem}
\label{thm:mainformal}
Let $k, n \in \N$. There is an algorithm $\Alg$ that takes as input a ``runtime'' parameter $k/2 \leq \ell \leq n/8$, a ``bias guess'' $\epshat \in (0,1)$, and a $k$-XOR instance $\Psi$ with constraint hypergraph $\cH$ with $m$ hyperedges, and in time $(n+m)^{O(\ell)}$ outputs a partition $\cH_{+} \sqcup \cH_{-} \sqcup \cH_{0}= \cH$ with the following guarantee. Suppose further that degree-$\ell$ sum-of-squares can refute instances drawn from $\LPN(\cH, \1^n,0)$, i.e., that with probability $1 - 1/mn$ over $\Phi \sim \LPN(\cH, 0)$, it holds that 
\begin{equation*}
\max_{\pE \in \SoS_{2\ell}} \pE[\Phi(x)] \leq \frac12+\frac{\gamma \heps }{720\log(mn)} \mper
\end{equation*}
Let $\eps \in (0,1)$ and $x^* \in \Fits^n$. Then, with probability $1 - m|\heps-\eps| - 1/\poly(n)$ over the draw of $\Psi = \{\cH, \{b_C\}_{C \in \cH}\} \sim \LPN(\cH, x^*, \eps)$, it holds that (1) for every $C \in \cH_{+}$, we have $b_C = x^*_C$, (2) for every $C \in \cH_{-}$, we have $b_C = -x^*_C$, and (3) $\abs{\cH_{0}} \leq \gamma|\cH|$.
\end{theorem}
We note that \Cref{thm:mainformal}, together with known SoS refutation algorithms (\Cref{fact:hkmrefutation}), implies the following corollary, which is the formal version of \cref{cor:main}.
\begin{corollary}\label{cor:formalmain}
    Let $2\leq k, n \in \N$. There is an algorithm $\Alg$ that takes as input a ``runtime parameter'' $k/2 \leq \ell \leq n$, a ``bias guess'' $\epshat \in (0,1)$, and a $k$-XOR instance $\Psi = (\cH, \{b_C\}_{C \in \cH})$ with constraint hypergraph $\cH$ of size $\abs{\cH} = m \leq n^{O(k)}$, and in time $n^{O(\ell)}$ outputs a partition $\cH_{+} \sqcup \cH_{-} \sqcup \cH_{0}$ of $\cH$ with the following guarantee. Suppose that the input $\Psi$ is drawn from $\LPN(\cH, x^*, \eps)$ and $\abs{\cH} = m \geq c_kn\left(\frac{n}{\ell}\right)^{k/2-1}\log^5(n) \gamma^{-4}\heps^{-4}$ for some universal $c_k$ depending only on $k$. Then, with probability $1 -  m\abs{\epshat - \eps} - 1/\poly(n)$, it holds that (1) for every $C \in \cH_{+}$, we have $b_C = x^*_C$, (2) for every $C \in \cH_{-}$, we have $b_C = -x^*_C$, and (3) $\cH_{0}$ is a set of size $\leq \gamma|\cH|$.
\end{corollary}
\begin{proof}
    By \Cref{fact:hkmrefutation}, for sufficiently large $c_k$, with probability $1-1/mn$ over $\Psi\sim\LPN(\cH,\1^n,0)$, we have $\pE[\Psi(x)]\leq \frac12+\frac{\heps\gamma}{720\log(mn)}$ for all degree-$2\ell$ pseudoexpectations $\pE$. The result then holds with probability $1-m|\heps-\eps|-1/\poly(n)$ by \Cref{thm:mainformal}.
\end{proof}
\begin{remark}
    When $k$ is even, the result of \Cref{fact:hkmrefutation} holds when $m \geq O(1)^k \cdot(\frac{n}{\ell})^{\frac{k}{2}} \ell \log_2(n) \cdot \delta^{-2}$, with a dependence of $\delta^{-2}$ instead of $\delta^{-4}$ (see Remark~4.2 of \cite{HsiehKM23}). Thus the result of \Cref{cor:formalmain} holds when $|\cH|=m\geq c_kn\left(\frac{n}{\ell}\right)^{k/2-1}\log^3(n) \gamma^{-2}\heps^{-2}$.
\end{remark}
\begin{corollary}
Let $2\leq k, n \in \N$. There is an algorithm $\Alg$ that takes as input a ``runtime parameter'' $k/2 \leq \ell \leq n$ and a $k$-XOR instance $\Psi = (\cH, \{b_C\}_{C \in \cH})$ with constraint hypergraph $\cH$ of size $\abs{\cH} = m \leq n^{k}$, and in time $n^{O(\ell)}$ outputs an assignment $x \in \Fits^n$ with the following guarantee. Suppose that the input $\Psi$ is drawn from $\LPN(\cH, x^*, \eps)$ and $\abs{\cH} = m \geq c_kn\left(\frac{n}{\ell}\right)^{k/2-1}\log^5(n) \gamma^{-4}\eps^{-4}$ for some universal $c_k$ depending only on $k$. Then, with probability $1 - o(1)$, it holds that $\Psi(x) \geq \Psi(x^*) - \gamma$.

Note that with probability $1 - o(1)$, we have $\Psi(x^*) \geq \frac12+\frac{\eps}{2} - o(1)$ and $\val(\Psi) \leq \frac12+\frac{\eps}{2} + o(1)$, so $\Psi(x) \geq \val(\Psi) -\gamma- o(1)$ with probability $1 - o(1)$.

\end{corollary}
\newcommand{\chii}{\cH^{(i)}}
\begin{proof}
    Let $c_k$ be the constant from \Cref{cor:formalmain} and for all $i=0, 1, \ldots, mn$, run the algorithm of \Cref{thm:mainformal} with $\heps=i/mn$, let $\chii_+\sqcup \chii_-\sqcup\chii_0$ be its output, and solve the linear system with equations $\{x_C=b_C\}_{C\in \chii_+}\cup\{x_C=-b_C\}_{C\in \chii_-}$ for $x^{(i)}\in \{\pm 1\}^n$. Output the value of $x^{(i)}$ maximizing $\Psi(x^{(i)})$.

    There exists an $i$ such that $|\heps-\eps|\leq 1/mn$, so by \Cref{cor:formalmain}, with probability $1-m\abs{\heps-\eps}-1/\poly(n)=1-o(1)$, we have $|\chii_0|\leq \gamma|m|$ and $x^*_C=b_C$ for all $C\in \chii_+$ and $x^*_C=-b_C$ for all $C\in \chii_-$, so the system has at least one solution $x^{(i)}$ satisfying $b_Cx^{(i)}_C=b_Cx^*_C$ for all $C\notin \chii_0$ and thus $\Psi(x^{(i)})\geq \Psi(x^*)-\gamma$.
\end{proof}

As explained in \cref{sec:overview}, the proof of \cref{thm:mainformal} follows a simple two step process. First, we use Sum-of-Squares semirandom $k$-XOR refutation certificates to construct a ``noisy prediction'' algorithm, and then we the sample splitting trick to split the input $\Psi$ into independent copies $\Psi^{(1)}, \dots, \Psi^{(t)}$ from a semirandom $k$-XOR distribution with smaller bias.

The noise predictor is captured by the first lemma below, and the splitting trick is captured by the second lemma.

\begin{restatable}[Noise prediction algorithm]{lemma}{predalg}
\label{lem:predalg}
Let $k, n \in \N$. There is a deterministic algorithm $\predict$ that takes as input a ``runtime parameter'' $k/2 \leq \ell \leq n/8$, a ``bias guess'' $\epshat \in (0,1)$, and a $k$-XOR instance $\Psi$ with constraint hypergraph $\cH$ on $n$ variables with $m$ hyperedges, and in time $(n+m)^{O(\ell)}$ outputs a function $\pred(\Psi, \cdot) \colon \cH \to \Fits$ with the following guarantee. Suppose further that degree-$\ell$ sum-of-squares can refute instances drawn from $\LPN(\cH, \1^n, 0)$, i.e., that with probability $1 - 1/mn$ over $\Phi \sim \LPN(\cH, \1^n, 0)$, it holds that 
\begin{equation*}
\max_{\pE \in \SoS_{2\ell}} \pE[\Phi(x)] \leq \frac12+\frac{\heps \gamma }{18} \mper
\end{equation*}
Then there is an $n^{O(\ell)}$-time randomized algorithm $\prune$ that takes in $\cH$ and outputs some $\cH' \subseteq \cH$ of size at least $(1 - \gamma)m$ such that with probability $1-1/n-m|\heps-\eps|$, for every $C \in \cH'$ and every $y \in \Fits^n$,
\begin{equation*}
\Pr_{\Psi  = (\cH, \{b_C\}_{C \in \cH}) \sim \LPN(\cH, y, \eps)}[\pred(\Psi, C) = y_C] \geq 2/3 -m|\heps-\eps| \mper
\end{equation*}
\end{restatable}

\begin{restatable}[Sample splitting]{lemma}{splitting}
\label{lem:splitting}
There is a randomized algorithm that takes as input a bit $b \in \Fits$, a parameter $\epshat \in (0,1)$ and $t \in \N$, and in time $O(2^t)$ outputs $b_1, \dots, b_t \in \Fits^t$ with the following guarantee. Suppose that $b = s r$ where $s \in \Fits$ is a secret bit and $r \sim \Bias(\eps)$. Then, the joint distribution of $(b_1, \dots, b_t)$, over the randomness of the algorithm and $r \sim \Bias(\eps)$, is $\abs{\epshat - \eps}$-close in TV distance to $t$ independent samples from $\Bias(s\eps/2t)$.
\end{restatable}

We postpone the proofs of \cref{lem:predalg,lem:splitting} to \cref{sec:predalg,sec:splitting}, respectively. Below, we finish the proof of \cref{thm:mainformal}.
\newcommand{\dtv}{d_{\text{TV}}}

\newcommand{\boc}{b^{(1)}_C}
\newcommand{\btc}{b^{(t)}_C}
\newcommand{\bic}{b^{(i)}_C}

\begin{proof}[Proof of \cref{thm:mainformal}]
Our algorithm works as follows:
\begin{itemize}
    \item Set $t\coloneqq 20\log(mn)$ and for all $C\in \cH$, run the algorithm of \Cref{lem:splitting} on $(b_C, \heps,t)$ to generate bits $\boc, \ldots, \btc$.
    \item For all $1\leq i\leq t$, let $\Psi_i\coloneqq(\cH, \{\bic\}_{C\in \cH})$ run the algorithm $\predict$ (\Cref{alg:predict}) from \Cref{lem:predalg} on $\ell,\heps/2t,\Psi_i$ and let $\pred_i:\cH\rightarrow \{\pm 1\}$ denote its output.
    \item Run the algorithm $\prune$ (\Cref{alg:prune}) from \Cref{lem:predalg} on $\cH$ and let $\cH'$ denote its output.
    \item For all $C\in \cH'$, let $y_C=\sgn(\sum_{i\in [t]}\pred_i(C))$.
    \item Set $\cH_+=\{C:y_C=b_C\}$, $\cH_-=\{C:y_C=-b_C\}$, and $\cH_0=\cH\setminus \cH'$. Output $\cH_{+} \sqcup \cH_{-} \sqcup \cH_{0}$.
\end{itemize}
By \Cref{lem:splitting}, note that $(\boc, \ldots, \btc)$ is $|\heps-\eps|$-close in total variation distance to $t$ independent samples from $\Bias(x^*\eps/2t)$, so the total variation distance of $(\Psi_1, \ldots, \Psi_t)$ and $t$ independent $\LPN(\cH,x^*,\eps/2t)$ is at most $ m|\heps-\eps|$. Therefore, by \Cref{lem:predalg}, Chernoff, and a union bound, with probability $1-m|\heps-\eps| - t/n$, we have $\Pr[\pred_i(C)=x^*_C]\geq \frac23-\frac{m|\heps-\eps|}{2t}\geq \frac{3}{5}$ for all $C\in \cH'$, $i\in [t]$ and $\Pr[y_C=\sgn(\sum_{i\in [t]}\pred_i(C))=x^*_C]$ for all $C\in \cH$ with probability $1-m|\heps-\eps|-2t/n$.
\end{proof}

\subsection{The noise prediction algorithm: proof of \cref{lem:predalg}}
\label{sec:predalg}
In this subsection, we prove \cref{lem:predalg}, restated below.

\predalg*

\begin{proof}
    First, we state the algorithm $\predict$ for generating the predictor $\pred:\cH\to\{\pm 1\}$.
    \begin{mdframed}
    \begin{algorithm}[Predict]\label{alg:predict}
    \mbox{}
    \begin{description}
        \item[Input:] A system of $k$-XOR equations $(\cH, \{b_C\}_{C \in \cH})$, where $\cH\seq[n]^k$, and a parameter $2k\leq\ell\leq n/8$.
        \item[Output:] A function $\pred:\cH\to \{\pm 1\}$ of predictions.
        \item[Operation:]\mbox{} 
                \begin{enumerate}[(1)]
            \item Sample uniform independent bits $r_1, \ldots, r_n\in \{\pm 1\}$.
            \item Set  $\zeta_C=b_C\prod_{i\in C}r_i$ for all $C\in \cH$ and solve the SDP $\pE\in\arg\max_{\pE\in\SoS_{2\ell}}\pE[\sum_{C\in\cH}\zeta_Cx_C]$.
            \item Let $\pred:\cH\rightarrow \{\pm 1\}$ where $\pred(C)\coloneqq \left(\prod_{i\in C}r_i\right) \sgn(\pE[x_C])$. Output $\pred$.

\end{enumerate}
    \end{description}

    \end{algorithm}
    \end{mdframed}
    Next, we state the algorithm $\prune$ for finding $\cH'$. 
    
    \begin{mdframed}
        \begin{algorithm}[Prune]\label{alg:prune}
    \mbox{}
    \begin{description}

        \item[Input:] A $k$-uniform hypergraph $\cH\subseteq [n]^k$
        \item[Output:] A subset $\cH'\subseteq \cH$.
        \item[Operation:]\mbox{} 
        \begin{enumerate}[(1)]

    \item For $1\leq i\leq T\coloneqq \max(1300\gamma^{-1}\log n, 288\log (mn))$ iterations, sample independent bits $\xii\sim \Bias(\heps)$, run $\predict$ on $(\cH,\{\xii_C\}_{C\in \cH})$, and let $\predi(C)$ be the result.
    \item For all $C\in \cH$, let $q(C)\coloneqq \#\{i:\predi(C)=-1\}$. Output $\cH'\coloneqq\{C\in \cH:q(C)\leq \frac{1}{4} T\}$.
        \end{enumerate}

    \end{description}
    \end{algorithm}
    \end{mdframed}
    \newcommand{\ri}{r^{(i)}}
    Run $\prune$ on $\cH$.
    For any $1\leq i\leq T$, let $\ri_1,\ldots, \ri_n$ and $\pei$ be the random bits and pseudoexpectation from the call to $\predict$. Note that $(\cH,\{\xii_C\}_{C\in \cH})\sim\LPN(\cH, \1^n,\heps)$. For any $C\in \cH$, we have $\predi(C)\neq x^*_C=1$ if $\sgn(\pei[x_C])\ri_C\neq 1$, so $\1\{\predi(C)\neq 1\}\leq 1-\ri_C\pei[x_C]$. Thus the total number of incorrect predictions for the $i$th iteration is upper bounded by $\sum_{C\in \cH}(1-\ri_C\pei[x_C])$. By optimality, we have $\sum_{C\in \cH}\xii_C\leq \sum_{C\in \cH}\xii_C \ri_C \pei[x_C]$, so
   \begin{align*}
       \heps\sum_{C\in \cH}(1-r_C\pE[x_C])
       &=\sum_{C\in \cH}(\xii_C-\heps)(\ri_C\pE[x_C]-1)+\sum_{C\in \cH}\xii_C(1-\ri_C\pei[x_C])
       \\&\leq \sum_{C\in \cH}(\xii_C-\heps)(\ri_C\pei[x_C]-1)
       \\&\leq \sup_{\pE'\in \SoS_{2\ell}}\sum_{C\in \cH}(\xii_C-\heps)(\pE'[x_C]-1)\mper
   \end{align*}
   For all $C\in \cH$, let $\xiii$ be an independent $\Bias(\epshat)$ and $\sigma_C$ be an independent Rademacher. Then
    \begin{align*}
        \E_\zi\sup_{\pE'\in \SoS_{2\ell}}\sum_{C\in \cH}(\xii_C-\heps)(\pE'[x_C]-1)
        &=\E_\xii\sup_{\pE'\in \SoS_{2\ell}}\sum_{C\in \cH}(\xii_C-\E \xiii_C)(\pE'[x_C]-1)
        \\&\leq \E_{\xii,\xiii}\sup_{\pE'\in \SoS_{2\ell}}\sum_{C\in \cH}(\xii_C-\xiii_C)(\pE'[x_C]-1)
        \\&\leq 2\E_{\sigma}\sup_{\pE'\in \SoS_{2\ell}}\sum_{C\in \cH}\sigma_C\pE'[x_C]+2\E_\sigma \sum_{C\in \cH}\sigma_C
        \\&\leq 2\frac{\heps\gamma m}{9}+ \frac{2}{n}\leq \frac{3\heps\gamma m}{13}\mcom
    \end{align*}
    where the last step follows from the refutation guarantee, so the expected number of mistakes is at most $3\gamma m/13$. By Chernoff, the total number of incorrect predictions over all $T$ rounds is at most $\gamma mT/4$ with probability $1-1/n$. Therefore, since $\sum_Cq(C)$ counts the total number of $i,C$ such that $\pred^{(i)}(C)=-1$ over all $T$ iterations, we have
\begin{align*}
    |\cH\setminus \cH'|\leq \frac{4}{T}\sum_{C\notin \cH'} q(C)\leq \frac{4}{T}\sum_{C}q(C)\leq \frac{4}{T}\cdot \frac{\gamma m}{4}T\leq\gamma|\cH|\mper
\end{align*}
Now we bound the prediction accuracy given a planted XOR instance with sufficiently close bias $\eps$. By Chernoff and a union bound, with probability $1-m\exp(-T/144)\geq 1-1/n$, we have $q(C)\geq \E[q(C)]- \frac{1}{12}T$. Then for all $C\in \cH'$, we have $\E q(C)\leq \frac{1}{4}T+\frac{1}{12}T=\frac{1}{3}T$. Now consider $\Psi = (\cH,\{b_C\}_{C\in \cH})\sim\LPN(\cH,y,\eps)$ and suppose we run $\predict$ on $(\Psi, \heps,\ell)$. Note that $\dtv(\LPN(\cH,y,\eps),\LPN(\cH,y,\heps))\leq |\heps-\eps|m$, and that $\Pr[\pred(C)=y_C]$ does not depend on $y$ as $\pred$ randomizes the plant to some $y\odot r$ before solving the SDP. Thus with probability $1-1/n-|\heps-\eps|m$, for all $C\in \cH'$,
\begin{align*}
    \Pr_{\Psi\sim\LPN(\cH,y,\eps)}[\pred(C)\neq y_C]
    &\leq \Pr_{\Psi\sim\LPN(\cH,y,\heps)}[\pred(C)\neq y_C]+m|\heps-\eps|
    \\&\leq \Pr_{\Psi\sim\LPN(\cH,\1^n,\heps)}[\pred(C)\neq 1]+m|\heps-\eps|
    \\&=\frac{1}{T}\E q(C)+m|\heps-\eps|\\&\leq \frac{1}{3}+m|\heps-\eps|
    \mper \qedhere
\end{align*}
\end{proof}
\subsection{The sample splitting algorithm: proof of \cref{lem:splitting}}
\label{sec:splitting}
In this subsection, we prove \cref{lem:splitting}, restated below
\splitting*
\begin{proof}
For all $y\in \{\pm 1\}^t$ and any $\alpha \in [0,1]$, let 
    $R^{(\pm)}_{\alpha}(y)\coloneqq\prod_{i\in [t]}\frac{1\pm y_i \alpha}{2}$ be the density functions of $t$ independent $\Bias(\alpha)$ and $\Bias(-\alpha)$ random variables, respectively. Given input $b$, we sample $(b_1, \ldots, b_t)=(y_1, \ldots, y_t)$ with probability
    \begin{align*}
        g^{(b)}(y_1, \ldots, y_t)\coloneqq \frac{1+b/\epshat}{2} R^{(+)}_{\epshat/2t}(y)+\frac{1-b/\epshat}{2}R^{(-)}_{\epshat/2t}(y)\mper
    \end{align*}
    First we show this is nonnegative for all $y\in \{\pm 1\}^t$. Note that $R^{(+)}_{\alpha}/R^{(-)}_{\alpha},R^{(-)}_{\alpha}/R^{(+)}_{\alpha}\geq (\frac{1-\alpha}{1+\alpha})^t\geq 1-2t\alpha$. Without loss of generality, assume $b=1$ (the same calculation holds with $R^{(+)}_{\epshat/2t}(y)$ in place of $R^{(-)}_{\epshat/2t}(y)$ when $b=-1$). Then
    \begin{align*}
        g^{(1)}(y_1, \ldots, y_t)
        \geq \left(\frac{1+1/\epshat}{2} (1-\heps)+\frac{1-1/\epshat}{2}\right)R^{(-)}_{\epshat/2t}(y)=\frac{1-\heps}{2}R^{(-)}_{\epshat/2t}(y)\geq 0\mper
    \end{align*}
    To show that each $b_i$ has the ideal bias distribution, let $g(y_1, \ldots, y_t)\coloneqq \E_b[g^{(b)}]$ and first note that 
    \begin{align*}
        \Pr[(b_1, \ldots, b_t)=(y_1, \ldots, y_t)]
        =g(y_1, \ldots, y_t)
        =\frac{1+s\eps/\epshat}{2}R^{(+)}_{\epshat/2t}(y)+\frac{1-s\eps/\epshat}{2}R^{(-)}_{\epshat/2t}(y)\mper
    \end{align*}
    For intuition, observe that since $b\sim\Bias(s\eps)$, we have  
    \begin{align*}
        \Pr[b_i=1]=\frac{1+s\eps/\epshat}{2}\cdot\frac{1+\epshat/2t}{2}+\frac{1-s\eps/\epshat}{2}\cdot\frac{1-\epshat/2t}{2}=\frac{1+s\eps/2t}{2}\mcom
    \end{align*}
    so each $b_i$ has distribution $\Bias(s\eps/2t)$.
    
    Now we will show the desired bound on the total variation. Let $h(y_1, \ldots, y_t)\coloneqq \prod_{i\in [t]}\frac{1+y_i s\eps/2t}{2}$ be the pdf of independent $\Bias(s\eps/2t)$ random variables. Observe that $h = R^{(s)}_{\eps/2t}$. By symmetry, we may assume that $s=1$.
    
    Then the total variation distance between the distribution $g$ and the distribution $h = R^{(+)}_{\eps/2t}$ is $\dtv(g, h) \leq \dtv(g, R^{(+)}_{\epshat/2t}) + \dtv(R^{(+)}_{\eps/2t}, R^{(+)}_{\epshat/2t})$. The latter term is at most $t \dtv(\Bias(\epshat/2t),\Bias(\eps/2t))$, which is $\frac{1}{2} \abs{\epshat - \eps}$.
    It remains to bound $\dtv(g, R^{(+)}_{\epshat/2t})$. We have
   
            \begin{align*}
	\dtv(g, R^{(+)}_{\epshat/2t}) 
    &= \frac{1}{2} \sum_{y\in \{\pm 1\}^t}|g(y)-R^{(+)}_{\epshat/2t}(y)| 
    \\&= \frac{1}{2}\sum_{y\in \{\pm 1\}^t}\left( \big| {\left(\frac{1+\eps/\epshat}{2}\right)R^{(+)}_{\epshat/2t}(y)+\left(\frac{1-\eps/\epshat}{2}\right)R^{(-)}_{\epshat/2t}(y) - R^{(+)}_{\epshat/2t}(y) \big| } \right)
    \\
	& = \frac{1}{2} \sum_{y\in \{\pm 1\}^t}|g(y)-R^{(+)}_{\epshat/2t}(y)| 
    \\&= \frac{1}{2} \left(\frac{\abs{1-\eps/\epshat}}{2}\right) \sum_{y\in \{\pm 1\}^t} \big| R^{(+)}_{\epshat/2t}(y) - R^{(-)}_{\epshat/2t}(y)  \big|  
    \\&=\left(\frac{\abs{\epshat-\eps}}{2 \epshat}\right) \dtv(R^{(+)}_{\epshat/2t}, R^{(-)}_{\epshat/2t} )\\
	& \leq \left(\frac{\abs{\epshat-\eps}}{2 \epshat}\right) \cdot t \dtv(\Bias(\epshat/2t), \Bias(-\epshat/2t)) 
    = \left(\frac{\abs{\epshat-\eps}}{2 \epshat}\right) \cdot t \cdot \frac{\epshat}{2t} =  \frac{\abs{\epshat-\eps}}{4} \mper 
    \end{align*}
    Hence, $\dtv(g, h) \leq \frac{3}{4} \abs{\epshat - \eps} <\abs{\epshat - \eps}$, as required.
\end{proof}
\begin{remark}
    We note that \cref{lem:splitting} is implied by the more general ``cloning theorems'', see~\cite[Lemma~14]{BrennanBH19} and~\cite[Lemma~20]{BrennanB20}. In fact, \cite[Lemma~20]{BrennanB20} can sample the independent bits $(b_1,\ldots,b_t)$ in $\poly(t)$ time (as opposed to the naive $O(2^t)$ time that it takes our algorithm, which writes down the entire PDF of $g^{(b)}$ to sample $(b_1, \ldots, b_t)$). 
\end{remark}

\section{Solving Semirandom Planted CSPs}
\label{sec:generalcsp}
In this section, we prove \cref{mthm:main2,cor:semirandomcsp}. We state the formal versions below.
\begin{theorem}\label{thm:formalmain2}
Let $k, n \in \N$. There is an algorithm $\Alg$ that takes as input a ``runtime parameter'' $k/2 \leq \ell \leq n$ and a $k$-CSP $\Psi$ with predicate $P$ and constraint hypergraph $\cH$ of size $\abs{\cH} = m \leq n^{O(k)}$, and in time $n^{O(\ell)}$ outputs an assignment $x \in \Fits^n$ with the following guarantee. For all $S\subseteq [k]$, let $\cH_S$ be the $|S|$-uniform hypergraph where $C_{|S}\in \cH_S$ for all $C\in \cH$. Suppose that for all $|S|\geq 2$ with probability $1 - o(1)$ over $\Phi \sim \LPN(\cH_S, \1^n, 0)$, degree-$\ell$ Sum-of-Squares can certify that $\val(\Phi) \leq \frac{1}{2} + \delta^2/(2^{2k+5}\cdot 720\log (mn))$, and either the same holds for $|S|=1$ or we have $m\geq \Omega(n\log n/\gamma\eps^2)$. Then, if $\Psi \sim \Psi(\cH, x^*, P, \cQ)$ for any $x^*$, with probability $1 -  o(1)$, it holds that $\Psi(x) \geq 1 - \delta$.
\end{theorem}
\begin{corollary}\label{cor:fullsemirandomcsp}
Let $2\leq k, n \in \N$. There is an algorithm $\Alg$ that takes as input a ``runtime parameter'' $k/2 \leq \ell \leq n$ and a $k$-CSP $\Psi$ with predicate $P$ and constraint hypergraph $\cH$ of size $\abs{\cH} = m \leq n^{O(k)}$, and in time $n^{O(\ell)}$ outputs an assignment $x \in \Fits^n$ with the following guarantee. Suppose that the input $\Psi$ is drawn from $\Psi(\cH, x^*, P, \cQ)$ and $\abs{\cH} = m \geq c_kn\left(\frac{n}{\ell}\right)^{k/2-1}\log^5 n\delta^{-8}$ for some universal constant $c_k$ depending only on $k$. Then, if $\Psi \sim \Psi(\cH, x^*, P, \cQ)$ for any $x^*$, with probability $1 - o(1)$, it holds that $\Psi(x) \geq 1 - \delta$.
\end{corollary}
\begin{proof}[Proof of \Cref{cor:fullsemirandomcsp}]
    By \Cref{fact:hkmrefutation}, for sufficiently large $c_k$, for any $S\subseteq [k]$ such that $|S|\geq 2$, with probability $1-1/mn$ over $\Psi\sim\LPN(\cH_S,\1^n,0)$ we have $\pE[\Psi(x)]\leq \frac12+\frac{\delta^2}{2^{2k+5}720\log(mn)}$ for all degree-$2\ell$ pseudoexpectations $\pE$. Since we have $m\geq \Omega(n/\gamma\eps^2)$, the result then holds with probability $1-m|\heps-\eps|-1/\poly(n)$ by \Cref{thm:formalmain2}.
\end{proof}\newcommand{\hatq}{\hat{Q}}
Recall the definition of a semirandom planted $k$-CSP $\Phi$ with predicate $P:\{\pm 1\}^k\rightarrow\{0,1\}$ and planting distribution $\cQ$ from \Cref{def:semirandomplantedcsp}. Let $(\hatq(S))_{S\subseteq [k]}$ be the Fourier coefficients of the distribution function of $\cQ$, i.e. $\cQ(y)\coloneqq \sum_{S\subseteq [k]}\hatq(S)y_S$.

Using $\Psi$, we define $|S|$-XOR instances for all $\emptyset\neq S\subseteq [k]$ as follows.
\begin{definition}
    Let $\Psi$ be a planted $k$-ary CSP with clause set $\cH\subseteq [n]^k$ and negations $b_C\in \{\pm 1\}^k$. For all $\emptyset\neq S\subseteq [k]$, define $\phisp,\phism$ to be the $|S|$-XOR instances where, for all $C\in \cH$, $\phisp$ contains the constraint $x_{C_S}= b_{C_S}$, where $x_{C_S}\coloneqq \prod_{i\in S}x_{C_i}$ and $b_{C_S}\coloneqq\prod_{i\in S}b_{C,i}$, and $\phism$ contains the constraint $x_{C_S}= -b_{C_S}$.
\end{definition}
We note that $\phisp,\phism$ are noisy planted $k$-XOR instances with the same planted assignment $x^*$.
\begin{fact}[Claim~4.2 of \cite{GuruswamiHKM23}]\label{fact:xordist}
    $\phisp\sim\LPN(\cH_S, x^*, 2^k\hatq(S))$ and $\phism\sim \LPN(\cH_S, x^*, -2^k\hatq(S))$
\end{fact}
\newcommand{\chsep}{\cH^{(S,\heps_S)}_+}
\newcommand{\chsem}{\cH^{(S,\heps_S)}_-}
\newcommand{\chsez}{\cH^{(S,\heps_S)}_0}
\newcommand{\xse}{x^{(S,\vec{\heps})}}
We modify the reduction in Section~4 of \cite{GuruswamiHKM23}, which first guesses which $S\subseteq [k]$ satisfy $|\hat{Q}(S)|\geq \eps/2^k$, then guesses $\sgn(\hat{Q}(S))$ for each of those $S$, and uses their noisy planted semirandom $k$-XOR algorithm to solve either $\phisp$ or $\phism$, depending on the sign guess. Since our more general noisy planted $k$-XOR algorithm requires us to guess $\heps$ such that $|\heps-2^k|\hatq(S)||\leq 1/\poly(mn)$ to achieve our soundness guarantee, we additionally guess a noise level $\heps_S$ for each $\phisp$ or $\phism$.

We first state two simple results shown as part of the proof of Theorem~2 of \cite{GuruswamiHKM23}. The first shows that for any clause $C\in \cH$ such that $\cQ$ has sufficiently large probability mass on $b_C\odot x^*_C$, then if we have some $x\in \{\pm 1\}$ consistent with $x^*$ on all subsets $S\subseteq [k]$ except those where $|\hatq(S)|$ is small, $x$ satisfies clause $C$. The second states that with high probability, all but $O_k(\eps)$ fraction of clauses $C$ will have a planted value with large enough probability mass. 
\begin{fact}[Step~2, Proof of Theorem~2 of \cite{GuruswamiHKM23}]\label{fact:satisfiespredicate}
    Suppose $C\in \cH$ such that $Q(b_C\odot x^*_C)> 2\eps$ and for all $S\subseteq [k]$ such that $|\hatq(S)|\geq \eps/2^k$, we have $x_{C_S}=x^*_{C_S}$. Then $P(b_{C_1}x_{C_1}, \ldots, b_{C_k}x_{C_k})=1$.
\end{fact}
\begin{fact}[Step~3, Proof of Theorem~2 of \cite{GuruswamiHKM23}]\label{fact:notmanybad}
    With probability $1-1/\poly(n)$, for all but at most a $2^{k+2}\eps$ fraction of $C\in \cH$, we have $Q(b_C\odot x^*_C)> 2\eps$.
\end{fact}
Finally, we state the following result for solving planted $1$-XOR instances, which we use to solve $\phisp$ or $\phism$ when $|S|=1$.
\begin{fact}[Lemma~C.1 of \cite{GuruswamiHKM23}]\label{fact:1xor}
    Let $n\in \N$. There is an algorithm $\Alg$ that takes as input a $1$-XOR instance $\Psi$ with constraint set $\cH$ of size $m\geq \Omega(n\log n/\gamma\eps^2)$ and in time $\poly(m,n)$ outputs a partition $\cH_{+} \sqcup \cH_{-} \sqcup \cH_{0}= \cH$ such that for any $x^* \in \Fits^n$, with probability $1 - 1/\poly(n)$ over the draw of $\Psi = \{\cH, \{b_C\}_{C \in \cH}\} \sim \LPN(\cH, x^*, \eps)$, it holds that (1) for every $C \in \cH_{+}$, we have $b_C = x^*_C$, (2) for every $C \in \cH_{-}$, we have $b_C = -x^*_C$, and (3) $\abs{\cH_{0}} \leq \gamma|\cH|$.
\end{fact}

\begin{proof}[Proof of \Cref{thm:formalmain2}]
    We iterate over all $3^{2^k}$ functions $f:2^{[k]}\setminus \{\emptyset\}\to \{0, +, -\}$ and do the following given some $f$, for all $S\in f^{-1}(+)$, construct $\phisp$, and for all $S\in f^{-1}(-)$, construct $\phism$. Then for all $\phisp$ and $\phism$, iterate over all $\heps_S+i/mn$ for $0\leq i\leq mn$ and apply the algorithm of \Cref{thm:mainformal} to obtain a partition $\chsep\sqcup\chsem\sqcup\chsez$. Then for each $f$, $\vec{\heps}=\{\heps_S\}_{S\in f^{-1}(\{+,-\})}$, solve the linear system
    \begin{align*}
        \cL\coloneqq\bigcup_{S\in f^{-1}(\{+\})} &\{x_{C_S}=b_{C_S}\}_{C\in \chsep}\cup \{x_{C_S}=-b_{C_S}\}_{C\in \chsem}
        \\&\cup \bigcup_{S\in f^{-1}(\{-\})} \{x_{C_S}=-b_{C_S}\}_{C\in \chsep}\cup \{x_{C_S}=b_{C_S}\}_{C\in \chsem}
    \end{align*}
     to obtain some $x^{(S,\vec{\heps})}\in \{\pm 1\}^n$. Finally, output the value of $\xse$ that maximizes $\Psi(\xse)$.
    
    Then there exists some $f$ such that $f(S)=+$ if and only if $\hatq(S)\geq \eps/2^k$, $f(S)=-$ if and only if $\hatq(S)\leq -\eps/2^k$, and $f(S)=0$ otherwise, and for each $S\in f^{-1}(\{+,-\})$, some $\heps_S$ such that $|\heps_S-2^k|\hatq(S)||\leq 1/mn$. Then for all such $S$ where $|S|\geq 2$ by \Cref{fact:xordist}, we apply the algorithm \Cref{thm:mainformal} with $\gamma=\delta/2^{k+1}$ and $\eps=\delta/2^{k+3}$ (so $\heps\geq \delta/2^{k+4}$), and when $|S|=1$, we either do the same or apply the algorithm of \Cref{fact:1xor}. A union bound shows that with probability $1-2^k(m/(mn)+1/\poly(n))=1-o(1)$, simultaneously for all $S\in f^{-1}(\{+,-\})$ we have $x^*_{C_S}=b_{C_S}f(S)$ for all $C\in \chsep$, $x^*_{C_S}=-b_{C_S}f(S)$ for all $C\in \chsem$, and $|\chsez|\leq 2^{-k-1}\delta|\cH|$. Letting $\cH_0=\bigcup_{S\in f^{-1}(\{+,-\})}\chsez$, we get $|\cH_0|\leq \frac12\delta |\cH|$, and for all $C\notin \cH_0$, $S\in f^{-1}(\{+,-\})$, we have $x^*_{C_S}=\xse_{C_S}$. Then by \Cref{fact:satisfiespredicate} and \Cref{fact:notmanybad}, with probability $1-o(1)$, $P(b_{C_1}\xse_{C_1},\ldots, b_{C_k}\xse_{C_k})$ for all but at most $\frac12\delta|\cH|$ of $C\notin \cH_0$. Thus $\Psi(\xse)\geq 1-\delta$ with probability $1-o(1)$, as desired.
\end{proof}

\section*{Acknowledgments}
We thank Pravesh K.\ Kothari, Jun-Ting ``Tim'' Hsieh, and Venkatesan Guruswami for numerous helpful discussions over the years on the semirandom planted CSP problem. We also thank Madhur Tulsiani and Tushant Mittal for pointing out the connection between \cref{lem:splitting} and the general theory of Bernoulli factories.

\section*{Statement of AI Use}
This statement is intended solely to describe the role of AI in the research process and should not be read as an endorsement of such systems or of the companies that produce them. In the production of this paper, we have used AI models for the following tasks.
\begin{flushleft}
\begin{tabular}{lc}
\textbf{Activity} & \textbf{AI used} \\
Asking the research question & [No] \\
Coming up with the approach & [No] \\
Proof development & [Yes] \\
Writing and exposition & [No] \\
Checking for bugs & [Yes] \\
Other supporting tasks & [Yes] \\
\end{tabular}
\end{flushleft}

\parhead{Details.} The problem of designing an ``\cite{RaghavendraRS17}''-style algorithm for semirandom planted CSPs dates back to~\cite{GuruswamiHKM23}. Over the course of our previous work~\cite{BasuHLM26}, we came up with the ``learning algorithm'' approach explained in \cref{sec:overview} in one of our attempts to design an algorithm for \emph{random} planted CSPs. However, we were unable finish the proof ourselves, as we did not see how to prove a statement like \cref{lem:splitting}. 

A few weeks after the release of GPT-6 Astra, A.L.\ attempted to tackle the ``\cite{RaghavendraRS17}''-style semirandom planted CSP problem with the help of Astra, and succeeded after a few prompts with vacuous human input. Notably, Astra was not provided with the ``learning algorithm'' approach, which it came up with independently from us. However, from our perspective the key contribution of Astra was \cref{lem:splitting}, as we already had formulated and proven all the other key lemmas in this paper, such as \cref{lem:predalg}, ourselves. A subsequent literature search using GPT-6 Astra revealed that \cref{lem:splitting} was already present in the literature~\cite{BrennanBH19,BrennanB20}, and it seems plausible that GPT-6 Astra got the idea of \cref{lem:splitting} from the aforementioned papers, which it then used without attribution.

We have verified the correctness and originality of all content, including references, by completely rewriting the AI's output and manually incorporating it into the paper. We have not made use of AI in the writing process, except to check for bugs after a near-final draft was produced and to do a literature search for \cref{lem:splitting}.
\begingroup
\small
\bibliographystyle{alphaurl}
\bibliography{references.bib}
\endgroup

\appendix

\end{document}